# Strong Conflict–Free Coloring of Intervals

Luisa Gargano and Adele A. Rescigno
Dipartimento di Informatica,
University of Salerno,
84084 Fisciano (SA), Italy

**Abstract**

We consider the $k$–strong conflict-free coloring of a set of points on a line with respect to a family of intervals: Each point on the line must be assigned a color so that the coloring has to be conflict-free, in the sense that in every interval $I$ there are at least $k$ colors each appearing exactly once in $I$.

In this paper, we present a polynomial algorithm for the general problem; the algorithm has an approximation factor $5 - \frac{2}{k}$ when $k \geq 2$ and approximation factor 2 for $k = 1$.

In the special case the family contains all the possible intervals on the given set of points, we show that a 2 approximation algorithm exists, for any $k \geq 1$.

## 1 Introduction

A coloring of the vertices of a hypergraph is said to be conflict–free if any hyperedge contains a vertex whose color is unique among those assigned to the vertices of the hyperedge.

**Definition 1. (CF Coloring)** *A conflict free vertex coloring of a hypergraph $H = (V, \mathcal{E})$ is a function $C : V \to \mathbb{N}$ such that for each $e \in \mathcal{E}$ there exists a vertex $v \in e$ such that $C(u) \neq C(v)$ for any $u \in e$ with $u \neq v$.*

Conflict–free coloring was first considered in [14]. It was motivated by a frequency assignment problem in cellular networks: each base station is assigned a certain frequency which is used to transmit data within a given (circular) region. Each client can scan frequencies in search of a base station within its range with good reception. Once such a base station is found, the client establishes a radio link with it, using a frequency not shared by any other station within its range. Given a set of base–stations the goal is to minimize the number of assigned frequencies. The problem can be modeled by means of a hypergraph whose vertices correspond to the base stations; then a hyperedge $e$ corresponds to a region in which each point is covered by all the base stations in $e$ and by none in $V \setminus e$. A conflict free coloring of such a hypergraph corresponds to an assignment of frequencies to the base stations, which enables any client to connect to one of them (holding the unique frequency in the client's area) without interfering with the other base stations.

CF-coloring also finds application in RFID (Radio Frequency Identification) networks. RFID allows a reader device to sense the presence of a nearby object by reading a tag attached to the object itself. To improve coverage, multiple RFID readers can be deployed in an area. However, two readers trying to access a tagged device simultaneously might cause mutual interference. It can be shown that CF-coloring of the readers can be used to assure that every possible tag will have a time slot and a single reader trying to access it in that time slot [23].

The notion of $k$-Strong CF coloring ($k$-SCF coloring), first introduced in [1], extends that of CF-coloring. A $k$-SCF coloring is a coloring that remains conflict-free after an arbitrary collection of $k - 1$ vertices is deleted from the set. Thus a $k$-SCF coloring for $k = 1$ is simply a standard CF coloring.

**Definition 2.** (*$k$-SCF Coloring*) *Let $H = (V, \mathcal{E})$ be a hypergraph and $k$ be a positive integer. A coloring $C : V \to \mathbb{N}$ is called a $k$–Strong Conflict–Free Coloring if for any $e \in \mathcal{E}$*

- *if $|e| \leq k$ then $C(u) \neq C(v)$ for each $u, v \in e$ with $u \neq v$;*
- *if $|e| > k$ then at least $k$ colors are unique in $e$, namely there exists $c_1, c_2, \ldots, c_k \in \mathbb{N}$ such that $|\{v \mid v \in e, \ C(v) = c_i\}| = 1$ , for $i = 1, \ldots, k$.*

*The goal is to minimize the size of the range of the $k$-SCF coloring function $C$.*
*We denote by $\chi_k^*(H)$ the smallest number of colors in any possible $k$-SCF coloring of $H$.*

In the context of cellular networks, this can be viewed as ensuring that for any client in an area covered by at least $k$ base stations, there always exists at least $k$ different frequencies with which the client can communicate without interference. This can be used both to serve up to $k$ clients at the same location as well as to deal with malfunctioning of some base stations. Analogously, in the context of RFID networks, a $k$–SCF coloring corresponds to a fault-tolerant activation protocol, that is, every tag can be read as long as at most $k - 1$ readers are broken.

## 1.1 Our Results: Strong CF Coloring of points with respect to intervals

Several authors recently focused on the special case of CF colorings of the family of all intervals on a line with $n$ points; here an interval is intended as intersecting at least one point on the line and two intervals are considered equivalent if they contain the same points.
Hence, the problem can be modeled as the CF coloring of the hypergraph

$$H_n = (V, \mathcal{I}) \text{ with } V = \{1, \ldots, n\} \text{ and } \mathcal{I} = \{\{i, i+1, \ldots, j\} \mid 1 \leq i \leq j \leq n\}, \qquad (1)$$

where each interval is represented by the set of consecutive points it contains. An example is given in Figure 1.
Conflict-free coloring for intervals models the assignment of frequencies in a chain of unit disks; this arises in approximately unidimensional networks as in the case of agents moving along a road. Moreover, it is important because it plays a role in the study of conflict-free coloring for more complicated cases, as for example in the general case of CF coloring of unit disks [14, 19].

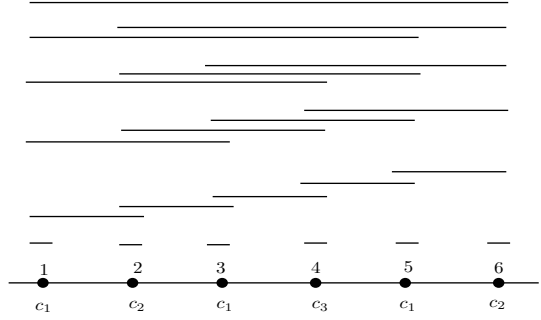

**Figure 1**: The hypergraph $H_6$ representing all the intervals on a line with 6 points. A 1-SCF coloring for $H_6$ is: $C(1) = C(3) = C(5) = c_1, C(2) = C(6) = c_2, C(4) = c_3$.

While some papers require the conflict-free property for all possible intervals on the line, in many applications good reception is needed only in some locations, in such a case it is necessary to supply only a given subset of the cells of the arrangement of the disks [17]. Indeed, in the context of channel assignment for broadcasting in wireless mesh network, it can occur that, at some step of the broadcasting process, sparse receivers of the broadcasted message are within the transmission range of a linear sequence of transmitters. In this case only part of the cells of the linear arrangements of disks representing the transmitters are involved [20, 24].

In this paper we consider the $k$–strong conflict–free coloring of points with respect to an arbitrary family of intervals. Hence, throughout the rest of the paper, we consider hypergraphs

$$H = (V, \mathcal{I}) \text{ with } V = \{1, \ldots, n\} \text{ and } \mathcal{I} \subseteq \{\{i, i+1, \ldots, j\} \mid 1 \leq i \leq j \leq n\}. \quad (2)$$

We shall refer to the above as *interval hypergraphs* and to $H_n$ as the *complete interval hypergraph*.

Figure 2 shows a 1-SCF coloring of the interval hypergraphs $H$ and $H'$ on 6 points. It is not hard to see that any 1-SCF coloring for $H$ needs at least 3 colors, while 2 colors suffice for $H'$.

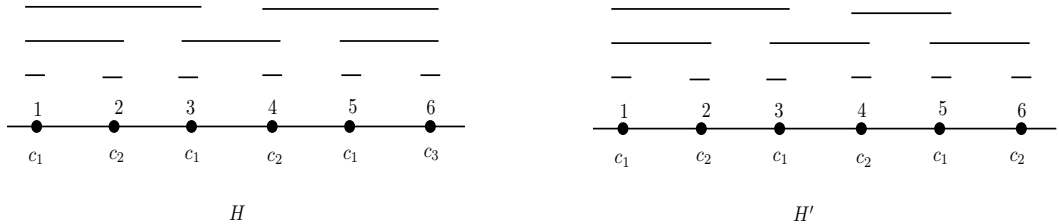

**Figure 2**: The hypergraphs $H$ and $H'$ on $V = \{1, \ldots, 6\}$ with their CF–colorings.

In Section 3, we give an algorithm which outputs a $k$–SCF coloring of the points of the input interval hypergraph $H$, for any fixed value of $k \geq 1$. We show the following results.

- The algorithm has an approximation factor $5 - 2/k$ in the case $k \geq 2$ (approximation factor 2 in the case $k = 1$);

- it optimally uses $k$ colors if for any $I, J \in \mathcal{I}$, interval $I$ is not a subset of $J$ and they differ in at least $k$ points.

In Section 4, we consider the problem of $k$–SCF coloring the complete interval hypergraph $H_n$. We give a very simple $k$–SCF coloring algorithm for $H_n$ that uses $k \left( \lfloor \log_2 \lceil \frac{n}{k} \rceil \rfloor + 1 \right)$ colors and show a lower bound of $\lceil \frac{k}{2} \rceil \lceil \log_2 \frac{n}{k} \rceil$ colors.

## 1.2 Related Work

Due to both its practical motivations and its theoretical interest, conflict–free colorings have attracted many researchers from the computer science community as well as from the mathematics community [2], [3], [4], [12], [14], [15], [16], [19], [21], [22]; a survey of results in the area is given in [23].

Conflict–free coloring in the case of the complete interval hypergraph has been first studied by Even et al. in [14]; they gave an algorithm showing that the problem can be optimally solved by using $\lfloor \log n \rfloor + 1$ colors.
The on–line version of the CF coloring problem for complete interval hypergraphs, where points arrive one by one and the coloring needs to remain CF all the time, has been subsequently considered in [5], [6], [7], [10], [11].

The problem of CF–coloring the points of a line with respect to an arbitrary family of intervals is studied in [17], [18], [8], [9]. In particular, [8] and [9] prove a 2-approximation algorithm for the problem. It is not known whether the CF–coloring problem is NP-hard for interval hypergraphs.

The $k$-SCF Coloring problem was first considered in [1] and has since then been studied in various papers under different scenarios, we refer the reader to [23] for more details on the subject. Recently, the minimum number of colors needed for $k$-SCF coloring the complete interval hypergraph $H_n$ has been studied in [13], where the exact number of needed colors for $k = 2$ and $k = 3$ has been determined. Moreover, Horev *et al.* show that $H_n$ admits a $k$-SCF coloring with $k \log n$ colors, for any $k$; this results follows as a specific case of a more general framework [16].

## 2 Notation

Through the rest of this paper we consider interval hypergraphs $H = (V, \mathcal{I})$, as defined in (2), on $n$ points, that is, $V = \{1, \ldots, n\}$.
Given $I \in \mathcal{I}$, we denote the *leftmost (e.g. minimum)* and the *rightmost (e.g. maximum)* of the points of the interval $I$ by

$$\ell(I) = \min\{p \mid p \in I\} \quad \text{and} \quad r(I) = \max\{p \mid p \in I\},$$

respectively; that is $I = \{\ell(I), \ell(I) + 1 \ldots, r(I)\}$.
We will use the following order relation on the intervals in the family $\mathcal{I}$.

**Definition 3. (Intervals ordering)**

- For all $I, J \in \mathcal{I}$
  $I \prec J \iff (r(I) < r(J))$ or $(r(I) = r(J)$ and $\ell(I) > \ell(J))$

- $I \in \mathcal{I}$ is called the $i$–th interval in $\mathcal{I}$ if $\mathcal{I} = \{I_1, \ldots, I_m\}$, $I_1 \prec I_2 \prec \ldots \prec I_m$, and $I = I_i$

Given a family $\mathcal{I}$, the subfamily of intervals of $\mathcal{I}$ that are not contained in $I$ and whose rightmost (resp. leftmost) point belongs to $I$ is denoted by $\mathcal{L}_\mathcal{I}(I)$ (resp. $\mathcal{R}_\mathcal{I}(I)$) that is

$$\mathcal{L}_\mathcal{I}(I) = \{J \in \mathcal{I} \mid J \nsubseteq I, \ r(J) \in I\} \text{ and } \mathcal{R}_\mathcal{I}(I) = \{J \in \mathcal{I} \mid J \nsubseteq I, \ \ell(J) \in I\}. \tag{3}$$

Clearly, $J \prec I$ (resp. $I \prec J$) for any $J \in \mathcal{L}_\mathcal{I}(I)$ (resp. $J \in \mathcal{R}_\mathcal{I}(I)$) with $J \neq I$.
Moreover, we denote by $\mathcal{M}_\mathcal{I}(I)$ the subfamily of all the intervals contained in $I \in \mathcal{I}$, that is

$$\mathcal{M}_\mathcal{I}(I) = \{J \mid J \in \mathcal{I}, \ J \subset I\}. \tag{4}$$

The index $\mathcal{I}$ is omitted whenever the family $\mathcal{I}$ is clear from the context.

Let $C : V \to \mathbb{N}$ the coloring function for $H = (V, \mathcal{I})$. We will use color 0 as a default color; that is initially all nodes are colored 0 and the algorithm will assign positive colors only to a subset of the points (alternatively, $C$ can be viewed as a partial coloring; this also models the scenario in which some base stations, in the cellular network, can be left inactive to save energy, [9], [18]):

- We say that an interval $I \in \mathcal{I}$ is $k$–**colored** if it contains at least $\min\{|I|, k\}$ unique colors, where a color $c \geq 1$ is unique in $I$ if there is exactly one point $p \in I$ such that $C(p) = c$.

## 3 A $k$-SCF Coloring Algorithm

In this section we present an algorithm for $k$-SCF coloring any hypergraph $H = (V, \mathcal{I})$ representing a family $\mathcal{I}$ of intervals on $V = \{1, \ldots, n\}$. We prove that our algorithm achieves an approximation ratio 2 if $k = 1$ and an approximation ratio $5 - \frac{2}{k}$ if $k \geq 2$; moreover, we show that the algorithm is optimal when $\mathcal{I}$ consists of intervals differing in at least $k$ points and not including any other interval in $\mathcal{I}$.

The $k$-SCF coloring algorithm, $k$-COLOR($\mathcal{I}$), is given in Fig. 3. The number of colors is upper bounded by the number of iterations performed by the algorithm times $c(k)$, where

$$c(k) = 2k + \left\lceil \frac{k}{2} \right\rceil - 1. \tag{5}$$

At each step $t$ of the algorithm a subset $P_t$ of points of $V$ is selected (through algorithm SELECT), then $c(k)$ colors are assigned in cyclic sequence to the ordered sequence (from the minimum to the maximum) of the selected points. The intervals that are $k$-colored at the end of the step $t$ are inserted in the set $\mathcal{X}_t$ and discarded. The algorithm ends when all the intervals in $\mathcal{I}$ have been discarded. At each step $t$ a new set of $c(k)$ colors is used.

Notice that a point $p \in V$ can be re–colored several times during different steps of the $k$–COLOR algorithm; its color at the end of the algorithm is the last assigned one.

The algorithm SELECT($\mathcal{I}_t$) considers intervals in $\mathcal{I}_t$ according to the $\prec$ relation and selects points so to have the required $\min\{|I|, k\}$ number of them in each interval. Namely, if $I$ is the $i$-th interval then it is considered at the $i$-th iteration of the **for** cycle; if less than $\min\{|I|, k\}$ points of $I$ have been already selected, then the algorithm adds the missing $\min\{|I|, k\} - |I \cap P_t|$ points of $I$ to $P_t$ (such points are the largest unselected ones of $I$).

$k$–**COLOR**($\mathcal{I}$)
Set $t = 1$.
$\mathcal{I}_1 = \mathcal{I}$.  *[$\mathcal{I}_t$ represents the set of intervals still to be $k$–colored at the beginning of step $t$]*
$\mathcal{X}_1 = \emptyset$.  *[$\mathcal{X}_t \subset \mathcal{I}_t$ contains the intervals that become $k$–colored during step $t$]*
**while** $\mathcal{I}_t \neq \emptyset$
    Execute the following **step** $t$
        **1.** Let $P_t = \{p_0, p_1, \ldots, p_{n_t}\}$ be the set of points returned by SELECT($\mathcal{I}_t$)
        **2.** **for** $i = 0$ **to** $n_t$
            Assign to $p_i$ color $c_i = (t-1)c(k) + (i \mod c(k)) + 1$
        **3.** **for** each $I \in \mathcal{I}_t$
            **if** $I$ is $k$-colored **then** $\mathcal{X}_t = \mathcal{X}_t \cup \{I\}$
        **4.** $\mathcal{I}_{t+1} = \mathcal{I}_t \setminus \mathcal{X}_t$
        **5.** $t = t + 1$

**SELECT**($\mathcal{I}_t$)
Set $P_t = \emptyset$.  *[$P_t$ represents the set of selected points at step $t$]*
**for** each $I \in \mathcal{I}_t$ by increasing order according to relation $\prec$ *[see Def.3]*
    **if** $|I \cap P_t| < \min\{|I|, k\}$ **then**
        **1.** Let $P_t(I)$ be the set containing the largest $\min\{|I|, k\} - |I \cap P_t|$ points of $I \setminus P_t$
        **2.** $P_t = P_t \cup P_t(I)$
Return $P_t$

**Fig. 3.** The $k$-SCF coloring algorithm for $H$.

**Example 1.** *Consider $H = (V, \mathcal{I})$, where $V = \{1, 2, \ldots, 23\}$ and $\mathcal{I}$ is the set of 13 intervals given in Fig. 4. Run $k$–COLOR($\mathcal{I}$) with $k = 2$; hence $c(2) = 4$ colors are used at each iteration. Initially, $\mathcal{I}_1 = \mathcal{I}$ and SELECT($\mathcal{I}_1$) returns $P_1 = \{3, 4, 7, 8, 9, 11, 12, 14, 15, 17, 18, 19, 20, 22, 23\}$ whose points are colored with $c_1, c_2, c_3, c_4$ in cyclic sequence. Only 3 intervals remain in $\mathcal{I}_2$; all the others are in $\mathcal{X}_1$, being 2-colored at the end of step 1. SELECT($\mathcal{I}_2$) returns $P_2 = \{14, 15, 23\}$ and these points are colored with $c_5, c_6, c_7$. Now $\mathcal{I}_3 = \mathcal{I}_2 \setminus \mathcal{X}_2 = \emptyset$ and the algorithm ends.*

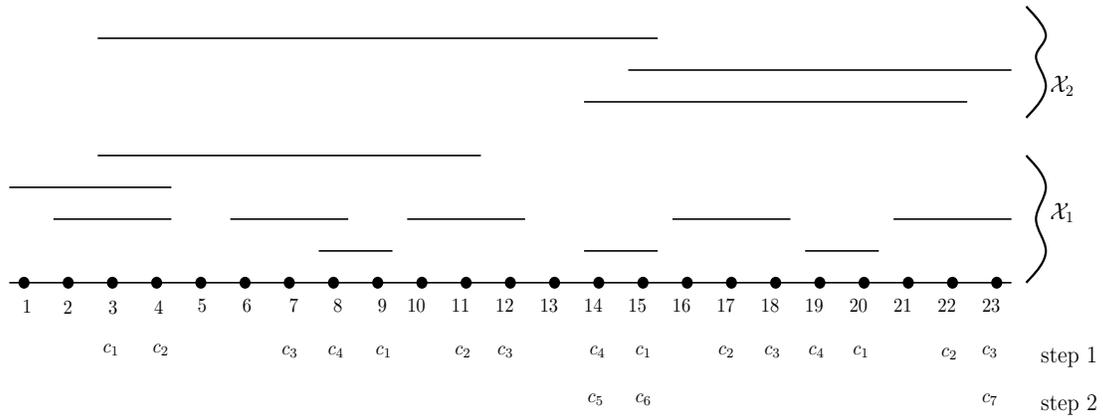

**Fig. 4**: Colors assigned by $k$–COLOR when $k = 2$.

We will prove the following theorem.

**Theorem 1.** *Given the interval hypergraph $H = (V, \mathcal{I})$, the algorithm $k$–COLOR($\mathcal{I}$) is a polynomial $k$–SCF coloring algorithm and uses $\chi(\mathcal{I}) < \frac{c(k)}{\lceil \frac{k}{2} \rceil} \chi_k^*(\mathcal{I})$ colors.*

## 3.1 Correctness of Algorithm $k$–COLOR

The following results characterize the intervals in the sets $\mathcal{X}_t$ produced by $k$–COLOR($\mathcal{I}$).

In the analysis of the algorithm SELECT($\mathcal{I}_t$) we will denote by $P_t$ only the output of the algorithm, that is, the set of all selected points. We denote by $P_t(i)$ the subset of points selected during iterations 1 up to $i$; that is, if $\mathcal{I}_t$ contains $I_1 \prec I_2 \prec, \ldots \prec I_{m_t}$, then

$$P_t(i) = P_t(I_1) \cup P_t(I_2) \ldots \cup P_t(I_i)$$

(cfr. line 1. of SELECT($\mathcal{I}_t$)), for all $i = 1, \ldots, m_t$.

**Lemma 1.** *Let $I \in \mathcal{I}_t$, $t \geq 1$.*
**a)** $|I \cap P_t| \geq \min\{|I|, k\}$;
**b)** *if $|I \cap P_t| \leq 4k - 2$ then $I \in \mathcal{X}_t$;*
**c)** $|I| \geq k$, *for each $t \geq 2$.*

**Proof.** Point a) follows since the algorithm SELECT($\mathcal{I}_t$) assures that $|I \cap P_t| \geq \min\{|I|, k\}$ are selected in each interval in $\mathcal{I}_t$.

We prove now b). Let the set of points selected by the algorithm SELECT($\mathcal{I}_t$) be $P_t = \{p_0, p_1, \cdots, p_{n_t}\}$ with $p_0 < p_1 < \ldots < p_{n_t}$. The step $t$ of the algorithm $k$-COLOR($\mathcal{I}$) assigns colors to the points in $P_t$ from the set $C = \{(t-1)c(k) + 1, (t-1)c(k) + 2, \cdots, tc(k)\}$ in cyclic sequence; namely point $p_i$ gets color $(t-1)c(k) + (i \mod c(k)) + 1$, for $i = 1, \ldots, n_t$.
Recalling that $c(k) = 2k + \lceil \frac{k}{2} \rceil - 1$, we have $4k - 2 \leq 2c(k) - k$. If $|I \cap P_t| \leq 4k - 2$, then at least $k \leq 2c(k) - |I \cap P_t|$ (or $|I \cap P_t|$ if less than $k$) among the colors assigned to the points in $I \cap P_t$ are unique. This, together with point a), assures that $I$ is contained in $\mathcal{X}_t$ and proves b).

Point c) is an immediate consequence of b) and of point a) with $t = 1$. $\square$

**Lemma 2.** *If $I \in \mathcal{X}_t$ then $\mathcal{M}_{\mathcal{I}_t}(I) \subseteq \mathcal{X}_t$.*

**Proof.** By contradiction, let $J \in \mathcal{I}_t \setminus \mathcal{X}_t$ and $J \subset I$. Since $J \in \mathcal{I}_t \setminus \mathcal{X}_t$ we have that, at the end of step $t$ of the algorithm $k$-COLOR($\mathcal{I}$), the interval $J$ is not $k$–colored. It follows that $|J \cap P_t| > 2c(k) - k$ must hold. Hence, $|I \cap P_t| \geq |J \cap P_t| > 2c(k) - k$ and $I$ also cannot be $k$–colored. This contradicts the hypothesis $I \in \mathcal{X}_t$. $\square$

**Lemma 3.** *If $\mathcal{M}_{\mathcal{I}_t}(I) = \emptyset$, then $|I \cap P_t| \leq 2k - 1$.*

**Proof.** Let $L \in \mathcal{I}_t$ be the last interval in $\mathcal{L}_{\mathcal{I}_t}(I)$, that is, $J \prec L$ for any $J \in \mathcal{L}_{\mathcal{I}_t}(I)$ with $J \neq L$. Assume $L$ is the $\ell$–th interval in $\mathcal{I}_t$. We first notice that $|L \cap I \cap P_t(\ell)| \leq k$. Otherwise, if $|L \cap I \cap P_t(\ell)| > k$ then there would exist an interval $K \subset L \cap I$ since SELECT($\mathcal{I}_t$) selects points only if they are useful to supply $k$ selected points to an interval; this is not possible since $\mathcal{M}_{\mathcal{I}_t}(I) = \emptyset$.

If $I \in \mathcal{I}_t$ then it is considered at iteration $\ell + 1$ and at most $k - |L \cap I \cap P_t(i)|$ are added to $P_t$. Hence

$$|I \cap P_t(i)| \leq k, \text{ for } i = \begin{cases} \ell & \text{if } I \notin \mathcal{I}_t, \\ \ell + 1 & \text{otherwise.} \end{cases} \quad (6)$$

Now, consider any $R \in \mathcal{R}_{\mathcal{I}_t}(I)$ and let $r$ be such that $R$ is the $r$-th interval in $\mathcal{I}_t$. Observe that $R \in \mathcal{R}_{\mathcal{I}_t}(I)$ implies that $R$ has at least one point outside $I$ and $r > i$.

If $P_t(r) \setminus P_t(i) \neq \emptyset$ then at least one of such points is selected at iteration $r$. This implies $|I \cap R \cap P_t(r)| \leq k - |(R \setminus I) \cap P_t| \leq k - 1$. This together with (6) gives the lemma.

If $P_t(r) \setminus P_t(i) = \emptyset$ then $|I \cap P_t| = |I \cap P_t(i)|$ and, by (6), the lemma holds. □

We prove now that $k$-COLOR($\mathcal{I}$) is a $k$–SCF coloring algorithm.

**Theorem 2.** *Given the interval hypergraph $H = (V, \mathcal{I})$, the algorithm $k$-COLOR($\mathcal{I}$) produces a $k$-SCF coloring of $H$.*

**Proof.** In order to prove the theorem, we show the following statement for each $t \geq 1$.

S($t$): *At the end of step $t$ of algorithm $k$-COLOR($\mathcal{I}$), each interval $I \in \bigcup_{i=1}^{t} \mathcal{X}_i$ is $k$–colored.*

The proof is by induction on $t$. For $t = 1$, the statement trivially follows by definition of $\mathcal{X}_1$. Consider now $t \geq 2$. Assume the statement be true for each $i \leq t - 1$. We prove that it holds for $t$. Notice that, by c) of Lemma 1, for any $I \in \mathcal{I}_t$ it holds $\min\{|I|, k\} = k$.

Clearly, if $I \in \mathcal{X}_t$, then $I$ is $k$–colored by definition of $\mathcal{X}_t$.

Consider then $I \in \mathcal{X}_i$ for some $i \leq t - 1$. By the inductive hypothesis $I$ is $k$–colored at the end of step $t - 1$. We will prove that, at the end of step $t$, the interval $I$ has $k$ unique colors, among $1, \ldots, t \cdot c(k)$. Indeed, by Lemma 2 we know that $\mathcal{M}_{\mathcal{I}_i}(I) \subseteq \mathcal{X}_i$; which implies that $\mathcal{M}_{\mathcal{I}_t}(I) = \emptyset$. Moreover, by Lemma 3, we have $|I \cap P_t| \leq 2k - 1 < c(k)$. This means that even if some points are recolored, all the assigned colors will be unique in $I$. □

## 3.2 Analysis of algorithm $k$-COLOR($\mathcal{I}$)

In this section we evaluate the approximation factor of the algorithm $k$-COLOR. We first give a lower bound tool (see also [13]).

**Theorem 3.** *Let $I_1, I_2, I \in \mathcal{I}$ with $I_1, I_2 \subset I$ and $I_1 \cap I_2 = \emptyset$. Let $\chi_1$ (resp. $\chi_2$) be the number of colors used by an optimal k–SCF coloring of $\mathcal{M}(I_1)$ (resp. $\mathcal{M}(I_2)$). Then the number of colors used by any optimal k–SCF coloring of $\mathcal{M}(I)$ is*

$$\chi^*(\mathcal{M}(I)) \geq \begin{cases} \max\{\chi_1, \chi_2\} & \text{if } k \leq |\chi_2 - \chi_1|, \\ \max\{\chi_1, \chi_2\} + \left\lceil \frac{k - |\chi_2 - \chi_1|}{2} \right\rceil & \text{otherwise.} \end{cases} \quad (7)$$

**Proof.** Suppose w.l.o.g. that $\chi_1 \leq \chi_2$. Recalling that $I_1, I_2 \in \mathcal{M}(I)$, we immediately get $\chi^*(\mathcal{M}(I)) \geq \chi_2$. Moreover, consider any coloring which colors the points in $I$ using $\chi_2 + c$ colors. Since $I_1 \cap I_2 = \emptyset$, we get that at most $(\chi_2 - \chi_1) + 2c$ colors can be unique in $I$. Hence, in order the coloring be $k$–SCF, it must hold $(\chi_2 - \chi_1) + 2c \geq k$. This immediately gives $c \geq \left\lceil \frac{k - |\chi_1 - \chi_2|}{2} \right\rceil$ and (7) holds. □

**Corollary 1.** *Let $I_1, I_2, I \in \mathcal{I}$ with $I_1, I_2 \subset I$ and $I_1 \cap I_2 = \emptyset$. If $\chi^*(\mathcal{M}(I_1)), \chi^*(\mathcal{M}(I_2)) \geq \chi$, then the number of colors used by any optimal k-SCF coloring of $\mathcal{M}(I)$ is*

$$\chi^*(\mathcal{M}(I)) \geq \chi + \left\lceil \frac{k}{2} \right\rceil. \quad (8)$$

In order to asses the approximation factor of the $k$–COLOR algorithm, we need the following result on the family $\mathcal{I}_t$ of intervals that still need to be $k$–colored after the step $t$ of the algorithm.

**Lemma 4.** *For each $I \in \mathcal{I}_t$, there exist at least two intervals $I', I'' \in \mathcal{I}_{t-1}$ such that $I', I'' \subset I$ and $I' \cap I'' = \emptyset$.*

**Proof.** We already know, by Lemma 3 and b) of Lemma 1, that if $\mathcal{M}_{\mathcal{I}_{t-1}}(I) = \emptyset$ then $I$ is $k$-colored at step $t$ and $I \notin \mathcal{I}_t$. Hence we can assume that $\mathcal{M}_{\mathcal{I}_{t-1}}(I) \neq \emptyset$ for any $I \in \mathcal{I}_t$.

To prove the lemma we proceed by contradiction. Suppose that there exists $I \in \mathcal{I}_t$ that does not include two disjoint intervals in $\mathcal{I}_{t-1}$, that is,

$$\bigcap_{K \in \mathcal{M}_{\mathcal{I}_{t-1}}(I)} K \neq \emptyset.$$

Let $K_1, K_2 \in \mathcal{M}_{\mathcal{I}_{t-1}}(I)$ be such that $K_1 \prec K' \prec K_2$ for each $K' \subset I$. Let $K_1$ and $K_2$ be the $(r_1 + 1)$–th and $r_2$–th interval in $\mathcal{I}_{t-1}$, respectively.

Reasoning as in Lemma 3, if we disregard the points added to $P_{t-1}$ at iterations $r_1+1, \ldots, r_2$ of SELECT($I_{t-1}$) when intervals in $\mathcal{M}_{\mathcal{I}_{t-1}}(I)$ are considered, then

$$\left| I \cap (P_{t-1} \setminus (P_{t-1}(r_2) \setminus P_{t-1}(r_1))) \right| \leq 2k - 1. \quad (9)$$

We want now to evaluate the number of points added to $P_{t-1}$ at any iteration of SELECT($I_{t-1}$) in which intervals $K_1, \ldots, K_2$ are considered, that is $|P_{t-1}(r_2) \setminus P_{t-1}(r_1)|$.

At most $k$ points are added to $P_{t-1}$ at iteration $r_1 + 1$, in order to complete $K_1$, that is

$$|P_t(r_1 + 1) - P_t(r_1)| \leq k.$$

Moreover at any iteration from $r_1 + 2$ to $r_2$ we only need to complete each interval up to $K_2$. Then the number of points added to $P_{t-1}$ will be

$$|P_{t-1}(r_2) - P_{t-1}(r_1 + 1)| \leq \begin{cases} k - 1 & \text{if } |P_t(r_1 + 1) - P_t(r_1)| > 0, \\ k & \text{if } |P_t(r_1 + 1) - P_t(r_1)| = 0. \end{cases}$$

In conclusion, the number of points added to $P_{t-1}$ when intervals $K_1, \ldots, K_2$ are considered is

$$|P_{t-1}(r_2) \setminus P_{t-1}(r_1)| \leq 2k - 1.$$

From this and (9) we have that

$$|I \cap P_{t-1}| \leq 4k - 2.$$

By Lemma 1, this implies that $I \in \mathcal{X}_{t-1}$ contradicting the hypothesis $I \in \mathcal{I}_t = \mathcal{I}_{t-1} \setminus \mathcal{X}_{t-1}$. □

In the following we assume that there exists at least an interval $I \in \mathcal{I}$ with $|I| \geq k$. Notice that if $|I| < k$ for each $I \in \mathcal{I}$, then each interval in $\mathcal{I}$ is $k$-colored after the first step of the algorithm $k$-COLOR($\mathcal{I}$) (even using for $c(k)$ the smaller value $\max\{|I| \mid I \in \mathcal{I}\}$).

**Lemma 5.** *Any $k$–SCF coloring algorithm on $\mathcal{I}_t$ uses at least $k + t\lceil\frac{k}{2}\rceil$ colors.*

**Proof.** By induction on $t$. For $t = 1$ the lemma is clearly true since any $k$-SCF coloring algorithm on $\mathcal{I}_1 = \mathcal{I}$ uses at least $k$ colors. Let $t \geq 2$ and assume that the lemma holds for any $i \leq t - 1$. Now, we prove it for $t$. By Lemma 4 and Corollary 1 we have that

$$\chi(\mathcal{I}_t) \geq \chi(\mathcal{I}_{t-1}) + \left\lceil\frac{k}{2}\right\rceil \geq k + (t-1)\left\lceil\frac{k}{2}\right\rceil + \left\lceil\frac{k}{2}\right\rceil = k + t\left\lceil\frac{k}{2}\right\rceil$$

where the last inequality is by the inductive hypothesis. □

We remark that the algorithm can be implemented in time $O(kn \log n)$, since in each step SELECT($\mathcal{I}_t$) can be implemented in $kn$ time (one does not actually need to separately consider all the intervals having the same right endpoint but only the $k$ shortest ones) and the number of steps is upper bounded by $O(\log n)$ – the worst case being the complete interval hypergraph. This together with the following Theorem 4 and Theorem 2, proves the desired Theorem 1.

**Theorem 4.** *Consider the interval hypergraph $H = (V, \mathcal{I})$. Then the total number of colors used by $k$-COLOR($\mathcal{I}$) is $\chi(\mathcal{I}) < \frac{c(k)}{\lceil\frac{k}{2}\rceil}\chi^*_k(\mathcal{I})$.*

**Proof.** Let $\delta$ be the last step of the algorithm $k$-COLOR($\mathcal{I}$). Then the total number of colors used by the algorithm is $\chi(\mathcal{I}) \leq \delta \cdot c(k)$; since each step of the algorithm uses a set of $c(k)$ colors. By Lemma 5, any optimal $k$–SCF coloring algorithm on $\mathcal{I}_\delta \subseteq \mathcal{I}$ uses at least $k + \delta \lceil \frac{k}{2} \rceil$ colors. This obviously implies $\chi^*(\mathcal{I}) \geq k + \delta \lceil \frac{k}{2} \rceil$. Hence,

$$\frac{\chi(\mathcal{I})}{\chi^*(\mathcal{I})} \leq \frac{\delta c(k)}{k + \delta \lceil \frac{k}{2} \rceil} < \frac{c(k)}{\lceil \frac{k}{2} \rceil}.$$

### 3.3 A Special Case

**Theorem 5.** *Let $H = (V, \mathcal{I})$. If for any $I, J \in \mathcal{I}$ such that $J \prec I$ and $I \cap J \neq \emptyset$ it holds*

  *a) $I \not\subseteq J$*   *b) $|I \setminus J| \geq k$*

*the algorithm $k$–COLOR($\mathcal{I}$), running with $c(k) = k$, optimally uses $k$ colors.*

**Proof.** Let $I$ be the $i$-th interval in $\mathcal{I}$. We will prove that at the end of the first step of $k$–COLOR($\mathcal{I}$) for each $I \in \mathcal{I} = \mathcal{I}_1$ it holds $|I \cap P_1| = \min\{|I|, k\}$. This will imply that if we run algorithm $k$–COLOR($\mathcal{I}$) with $c(k) = k$ then each interval in $\mathcal{I}$ is optimally $k$-colored.

Indeed, by a) of Lemma 1, we know that $|I \cap P_1(i)| \geq \min\{|I|, k\}$. Moreover, as in the proof of Lemma 3, we can prove $|I \cap P_1(i)| \leq \min\{|I|, k\}$.

Now, consider any $R \in \mathcal{R}(I)$ and let $R$ be the $r$-th interval in $\mathcal{I}$. Since $|R \setminus I| \geq k \geq \min\{|R|, k\} - |R \cap P_1(r-1)|$ we have $|I \cap P_1(r)| \leq \min\{|I|, k\}$. □

## 4 A $k$–SCF coloring algorithm for $H_n$

In this section we present a $k$-SCF-coloring algorithm for the complete interval hypergraph $H_n = (V, \mathcal{I})$ with $V = \{1, \ldots, n\}$ and $\mathcal{I} = \{\{i, i+1, \ldots, j\} \mid 1 \leq i \leq j \leq n\}$. When $k = 1$ the algorithm reduces to the one in [14]. We assume that $n = hk$ for some integer $h \geq 1$. If $(h-1)k < n < hk$ then we can add the points $n+1, n+2, \ldots, hk$.

A simple $k$-SCF-coloring algorithm for $H_n$ can be obtained by partitioning the $n = hk$ points of $V$ in blocks $B(1), B(2), \cdots, B(h)$ of $k$ points and coloring their points recursively with the colors in the sets $C_1, \cdots, C_{\lfloor \log_2 h \rfloor + 1}$ where, for $1 \leq t \leq \lfloor \log_2 h \rfloor + 1$,

$$C_t = \{k(t-1) + 1, \cdots, kt\}.$$

In particular, the points in the median block $B(\lfloor \frac{h+1}{2} \rfloor)$ are colored with colors in $C_1$, then the points in the blocks $B(1), \cdots, B(\lfloor \frac{h+1}{2} \rfloor - 1)$ and in the blocks $B(\lfloor \frac{h+1}{2} \rfloor + 1), \cdots, B(h)$ are recursively colored with the same colors in the sets $C_2, \cdots, C_{\lfloor \log_2 h \rfloor + 1}$.

Formally the algorithm is given in Fig.5. It starts calling $(k, n)$–COLOR$(1, h, 1)$.

The proof that the algorithm $(k, n)$–COLOR$(1, h, 1)$ provides a $k$-SCF coloring for $H_n$ can be easily derived by that presented in [14, 23]. Furthermore, since at each of the $\lfloor \log_2 h \rfloor + 1$ recursive steps of algorithm $(k,n)$–COLOR a new set of $k$ colors is used, we have that the number of colors is at most $k(\lfloor \log_2 h \rfloor + 1)$. Hence, we get the following result.

**Lemma 6.** *At the end of algorithm $(k,n)$–COLOR$(1, \lceil \frac{n}{k} \rceil, 1)$ each $I \in \mathcal{I}$ is $k$-SCF colored and the number of used colors is at most $k\left(\lfloor \log_2 \lceil \frac{n}{k} \rceil \rfloor + 1\right)$.*

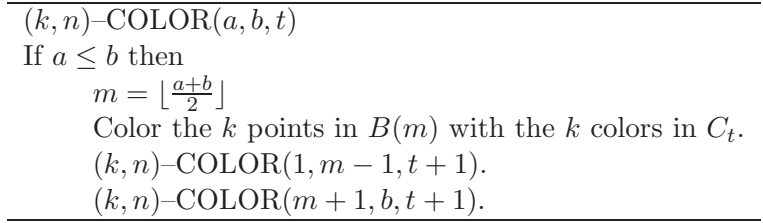

**Fig. 5.** $k$-SCF-coloring for $H_n$.

We remark that [16] shows that $\chi^*(H_n) \leq k \log n$ (as a specific case of a more general framework); however, we present the $(k,n)$–COLOR algorithm since it is very simple and gives a slightly better bound.

By Corollary 1 and considering that, for the complete interval hypergraph $H_n$, for each $I \in \mathcal{I}$, any of its subintervals $I' \subset I$ also belongs to $\mathcal{I}$, we get the following lower bound on $\chi^*(H_n)$.

**Corollary 2.** $\chi^*(H_n) \geq \lceil \frac{k}{2} \rceil \lceil \log_2 \frac{n}{k} \rceil$.

Lemma 6 together with Corollary 2 proves the following theorem.

**Theorem 6.** *The problem of $k$–SCF coloring the complete interval hypergraph $H_n = (V, \mathcal{I})$ admits a 2-approximation algorithm.*

## 5 Conclusions

We have presented an algorithm for $k$–SCF coloring an input interval hypergraph $H = (V, \mathcal{I})$, for any fixed value of $k \geq 1$. The algorithm has an approximation factor $5 - 2/k$ in the case $k \geq 2$, and an approximation factor 2 in the case $k = 1$. Furthermore, we have shown that the algorithm optimally uses $k$ colors if for any $I, J \in \mathcal{I}$, interval $I$ is not contained in $J$ and they differ for at least $k$ points. Finally, we have presented an algorithm for the complete interval hypergraph $H_n$ and we have proved that the algorithm has an approximation factor 2.

Several problems remains open in the area of SCF coloring of interval hypergraphs. Mainly the complexity of the CF-coloring problem has not been assessed. Is it possible to improve the approximation factor for SCF-coloring algorithms? Actually, it is not known whether the problem is NP-complete for general interval hypergraphs even for $k = 1$.
In case of the complete interval hypergraph $H_n$, it would be interesting to close the gap between the upper and lower bounds; in particular, is it possible to have a CF-coloring algorithm that improves the factor $k$ in Lemma 6 to the factor $\lceil \frac{k}{2} \rceil$?